\documentclass{pasj00}

\begin{document}
\SetRunningHead{Kochanov et al.}{Enhanced solar imaging with
SSRT}

\title{Imaging of the solar atmosphere by the Siberian Solar Radio Telescope
at 5.7 GHz with an enhanced dynamic range}

\author{Alexey \textsc{Kochanov},
Sergey \textsc{Anfinogentov},
Dmitry \textsc{Prosovetsky}, George
\textsc{Rudenko}, Victor \textsc{Grechnev}}
 \affil{Institute of
Solar-Terrestrial Physics SB RAS, Lermontov St. 126A, Irkutsk
664033, Russia}\email{anfinogentov@iszf.irk.ru,
kochanov@iszf.irk.ru}



\maketitle

\begin{abstract}
The Siberian Solar Radio Telescope (SSRT) is a solar-dedicated
directly-imaging interferometer observing the Sun at 5.7 GHz. The
SSRT operates in the two-dimensional mode since 1996. The imaging
principle of the SSRT restricts its opportunities in observations of
very bright flare sources, while it is possible to use `dirty'
images in studies of low-brightness features, which do not overlap
with side lobes from bright sources. The interactive CLEAN technique
routinely used for the SSRT data provides imaging of active regions
but consumes much time and efforts and does not reveal
low-brightness features below the CLEAN threshold. The newly
developed technique combines the CLEAN routine with the
directly-imaging capability of the SSRT and provides clean images
with an enhanced dynamic range automatically. These elaborations
considerably extend the range of tasks, which can be solved with the
SSRT. We show some examples of the present opportunities of the SSRT
and compare its data with the images produced by the Nobeyama
Radioheliograph at 17 GHz as well as observations in different
spectral ranges.
\end{abstract}

\section{Introduction}
 \label{sect:introduction}

Microwaves provide extensive opportunities to study quiet and
dynamic solar features from chromospheric to coronal heights by
observing their thermal free-free and gyromagnetic emissions. The
Siberian Solar Radio Telescope (SSRT;
\cite{Smolkov1986,Grechnev2003}) is a solar-dedicated interferometer
observing the Sun in total intensity (Stokes $I$) and circular
polarization (Stokes $V$) at a frequency of 5.7 GHz ($\lambda =
5.2$~cm). The SSRT is a rather old instrument initially proposed in
1960-s and constructed in 1970--1980-s. Having started to observe in
1980-s in the one-dimensional mode, the SSRT underwent a number of
upgrades of both the hardware and software systems.First
two-dimensional maps were obtained in 1991 from one-dimensional SSRT
observations by \citet{Alissan1992} using the Earth-rotation
aperture synthesis techniques. Fast one-dimensional imaging started
in 1992 \citep{Altyntsev1994}. Regular production of several images
per day in the two-dimensional mode commenced in 1996. The
observations at the SSRT have been interrupted in July 2013 due to a
fundamental upgrade.

\begin{figure*}
  \begin{center}
    \FigureFile(170mm,70mm){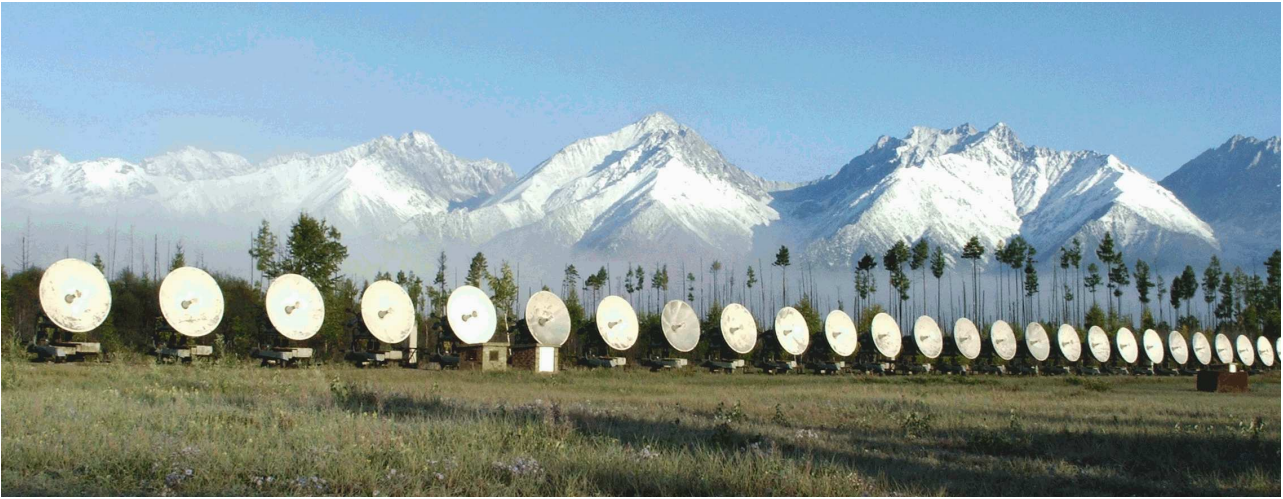}
  \end{center}
  \caption{View of the SSRT North arm.}
  \label{fig:ssrt_view}
 \end{figure*}

The SSRT is located in Badary forest area 220~km from Irkutsk at
N~$51^{\circ }45^{\prime }$, E~$102^{\circ }13^{\prime }$. The SSRT
is a cross-shaped interferometer consisting of two equidistant
linear arrays in the East--West and North--South directions, each of
128 antennas spaced by 4.9~m (Figure~\ref{fig:ssrt_view}). Each
linear interferometer has a baseline of 622.3~m. Unlike synthesizing
interferometers such as the Nobeyama Radioheliograph (NoRH;
\cite{Nakajima1994}), the SSRT produces images directly without
measuring visibilities. The imaging process invokes the solar
rotation for the scanning in the hour angle and the so-called
`frequency scanning' technique in the altitude (over 5.67--5.79
GHz). One full-disk solar image with a field of view of $42^{\prime}
\times 42^{\prime}$ is composed typically in 3 to 5 minutes. The
SSRT produces several tens of single-pass images, depending on the
Sun's declination (i.e., season). The overall SSRT sensitivity in
solar observations is about 1\,500~K for an image formed in a single
pass of the Sun (integration time 0.336~s). The system noise
contribution is about 800~K. The dynamic range of the raw data is
about 30 dB.

The directly-imaging operating principle of the SSRT determines a
relatively low level of the side lobes (22\% of the main maximum for
the non-distorted beam) and therefore allows one to use even `dirty'
maps occasionally, when radio sources on the Sun are not very bright
(\cite{Uralov1998};
\authorcite{Grechnev2003} \yearcite{Grechnev2003,
Grechnev2006quiet}). However, in a typical situation, contributions
from the beam's side lobes make the deconvolution of the SSRT images
quite necessary.

The basic deconvolution algorithm in the solar radio astronomy is
the CLEAN method proposed by \citet{Hogbom1974} due to its
robustness, although successful examples are known of invoking
different techniques such as MEM (e.g.,
\cite{Enome1995,Nindos1999}). CLEAN provides reliable results for
compact features like gyromagnetic sources in solar active regions
but is not well-suited for extended sources. The deconvolution of
such sources using the classical CLEAN method results in the
so-called `fringing', i.e., dividing a single continuous object into
a multitude of small bright features. These artifacts of the
original CLEAN method make it less suitable for the deconvolution of
extended solar features such as the `bald' solar disk, filaments,
coronal holes, etc. \citet{Nindos1999} showed that different
deconvolution methods such as the Steer CLEAN \citep{Steer1984} and
even MEM also suffered from this problem.

The performance of the CLEAN algorithm for extended sources can be
improved by adopting the multi-scale approach, in which the sources
to be recovered are modeled as being composed of various different
scale sizes. The deconvolution technique routinely used at the SSRT
up to the recent time was, in fact, one of the first implementations
of this approach \citep{Lesovoy2002}. This technique successfully
deconvolved bright sources in active regions, but did not recover
low-brightness features, thus decreasing the dynamic range of the
clean maps to about 20 dB. This circumstance is consistent with the
conclusion drawn by \citet{Koshiishi1994} for NoRH that the final
image quality is not purely determined by the hardware capabilities
of the radio heliograph being significantly dependent on the
image-restoration procedure.

For this reason, it was very difficult to carry out from
two-dimensional SSRT images systematic studies of low-brightness
features such as the center-to-limb brightness distribution,
supergranulation network, filaments and prominences, coronal holes,
etc. The first problem was due to the fact that the residual map
below the final CLEAN level (threshold) is usually not processed but
substituted with a model disk, so that all low-brightness features
disappear. A satisfactory way to recover the residual map was not
developed. On the other hand, usage of dirty maps requires a lot of
manual work to select appropriate images and can only be done in
exceptional cases. This way cannot be used systematically, e.g., to
enhance the sensitivity by the averaging of several images.

To overcome these limitations and actualize the potential dynamic
range of the SSRT, we have implemented a new imaging technique based
on a multi-scale CLEAN (MS-CLEAN) algorithm described by
\citet{Cornwell2008} and adjusted to the SSRT data with a recovery
of the residual map. This technique combines the MS-CLEAN routine
with the directly-imaging capability of SSRT and produces
enhanced-quality clean images automatically. From these images, it
has become possible to study low-brightness solar features.

The operating frequency of the SSRT (5.7 GHz) is three-times lower
than the lowest one of NoRH (17~GHz) that results in considerably
larger opacities of microwave sources observed by the SSRT with
respect to 17~GHz ones to both the thermal free-free and
gyromagnetic emissions. This circumstance determines a vast variety
of contributions from chromospheric to coronal features in SSRT
images.

Microwave observations of the quiet Sun at different frequencies
have been reviewed by \citet{Kundu1982} and more recently by
\citet{Shibasaki2011}. Microwave observations also offer wide
opportunities to study active low-brightness features such as
eruptive prominences and filaments from SSRT and NoRH images, as has
been demonstrated by, e.g.,
\citet{Hanaoka1994fil,Uralov2002,Shimojo2006};
\authorcite{Grechnev2006erupt}
(\yearcite{Grechnev2003,Grechnev2006quiet,Grechnev2006erupt}) and
others.

Among the issues, which can be addressed by the analysis of the
enhanced-quality SSRT images, there are the following. The
brightness distribution over the quiet Sun depends on the parameters
of the solar atmosphere and is promising to elaborate the models.
This distribution and its solar cycle variations at 17 GHz have been
studied in detail from NoRH data (e.g., \cite{Nindos1999};
\authorcite{Selhorst2004}
\yearcite{Selhorst2004,Selhorst2011}); however, only occasional
measurements in longer microwaves have been known (e.g.,
\cite{Kundu1982,Krissinel2005}), whose results do not complete
correspond to each other.

The microwave response to the supergranulation network has been
first detected at 2--4 cm in one-dimensional observations with the
RATAN-600 radio telescope \citep{Bogod1975, Bogod1978}. First
two-dimensional microwave observations of the network with the
Westerbork Synthesis Radio Telescope (WSRT) at a close (to SSRT)
wavelength of 6~cm have been reported by \citet{Kundu1979}.
Comparison of the microwave observations with the optical ones in
the Ca$^{+}$~K line has shown their correspondence. \citet{Benz1997}
compared observations of the network with the Very Large Array (VLA)
in microwaves at of 1.3, 2.0, and 3.6~cm with magnetic network
elements and soft X-ray images produced by Yohkoh/SXT. The authors
found the highest correspondence with magnetic network elements at
the shortest wavelengths that decreased at 3.6~cm and was lowest in
soft X-rays. There were very few similar observations with
non-solar-dedicated WSRT and VLA telescopes, so that the data they
provided are unique. The routine SSRT data appear to be promising to
study the `microwave network' as well.

In this paper we discuss the technique to routinely produce solar
images at the SSRT and its elaboration. We present some beacon
results on low-brightness features obtained in the analysis of
enhanced-quality images to illustrate the opportunities of the SSRT.
Elaboration of these results promises important contribution to the
solar physics. We also compare the SSRT data with images produced by
the Nobeyama Radioheliograph at 17 GHz as well as observations in
different spectral ranges.

Section~\ref{sect:imaging_techniques} outlines the imaging technique
routinely used at the SSRT to deconvolve bright sources in active
regions and describes our elaborated technique to produce the
enhanced-quality images. Section~\ref{sect:synoptic} briefly
describes the techniques to construct the microwave synoptic maps
and to supplement the microwave images with the magnetic field data.
Section~\ref{sect:features_in_ssrt_images} outlines observations of
some low-brightness features. Section~\ref{sect:summary} presents a
brief summary and outlines the long-standing problems and the plans.

\section{SSRT Imaging Techniques}
 \label{sect:imaging_techniques}

We consider the techniques to produce deconvolved images from
`dirty' SSRT data without an emphasis on the formation of the solar
images with the SSRT. A description of the latter issue can be found
in \citet{Smolkov1986} and \citet{Grechnev2003}.

\subsection{Imaging of Bright Sources in Active Regions}
The SSRT produces images directly without measuring visibilities. 
The interactive CLEAN technique~\citep{Lesovoy2002} routinely used for the 
SSRT data for more than one decade provides imaging of active regions in 
semi-automatic mode, consuming time and efforts.
The corresponding viewable SSRT images as well as digital data in 
FITS format (one pair of single-pass images per day) are available at
http:/\negthinspace/www.ssrt.org.ru/ and
ftp:/\negthinspace/iszf.irk.ru/pub/ssrt.

\subsection{Advanced Imaging Technique}
To achieve a fully automatic imaging at the SSRT with enhanced
opportunities capable of reproducing low-brightness features, a
hybrid approach was developed \citep{Kochanov2011}. The
deconvolution is based on a multi-scale version of the CLEAN
algorithm \citep{Cornwell2008}. The MS-CLEAN has been adjusted for
the SSRT imaging and supplemented with an additional adaptive
filtering to suppress residual noises and errors of the MS-CLEAN
algorithm. The major feature of the new technique is processing of
the residual map to recover low-brightness objects.

\subsubsection{Multi-Scale CLEAN for SSRT}

We have found the MS-CLEAN algorithm to be most suitable to produce
`clean' images with the SSRT. While implementing the MS-CLEAN,
selection of a proper size is crucial for each spatial scale. If the
selected scale is too small, then false sources appear in a `clean'
map (`fringing' effect). If the selected scale is too large, then
the source in the `clean' map becomes excessively smoothed, and a
spurious dark halo appears in the residual.

We have made several modifications of the original MS-CLEAN
algorithm to adapt it to the SSRT imaging. Below we describe our
algorithm using the notations according to \citet{Cornwell2008} and
omitting some details, which can be found in this paper.

\begin{itemize}
\item Initialize
   \begin{itemize}
       \item Model Image: $I^m=0$
       \item Residual image: $I^R =$ dirty image
       \item For each scale $\alpha$
           \begin{itemize}
               \item[$\circ$] Calculate the `clean' beam $m_c(\alpha)=H(r)\left[1-\left(\frac{r}{\alpha}\right)^2\right]$, where $r$ is the radial distance and $H(r)$ is a Hanning window function of $2 \alpha$ width centered at $r=0$
               \item[$\circ$] Calculate the scale-convolved `dirty' beam
               $m(\alpha)=\mathrm{PSF}*m_c(\alpha)$ (`$*$' denotes convolution)
               \item[$\circ$] Calculate the scale-convolved residual as $I^R_\alpha=I^R * m(\alpha)$.
               \item[$\circ$] Calculate the scale normalization coefficient $S(\alpha)=1/\sqrt{\sum\limits_{x,y}{m(\alpha)^2}} $
               \item[$\circ$] For each pair of scales $(\alpha_p, \alpha_q)$ calculate the cross term $B*m(\alpha_p)*m(\alpha_q)$
           \end{itemize}
   \end{itemize}
\item Repeat
   \begin{itemize}
   \item Find the position of the maximum on the residual map $I^R$ (modified)
   \item Choose the scale with a maximum residual value at the selected position after multiplying by the scale-dependent normalization coefficient $S(\alpha)$
   \item Add this component scaled by the loop gain and convolved with the `clean' beam of the selected scale to the current model.
   \item Update all residual images using the pre-computed terms.
   \end{itemize}
\item Until
   \begin{itemize}
   \item Either maximum iteration number reached
   \item Or $\max(I^r)<(\mathrm{CLEAN\ threshold})$
   \end{itemize}
\item Finalize
   \begin{itemize}
   \item add the residual $I^R$ to the model $I^m$ to get the restored image
   \end{itemize}
\end{itemize}

To prevent selecting the largest scale in all iterations, the
original MS-CLEAN algorithm uses the scale biases:
    \begin{equation} \label{eq:scale_bias}
    S(\alpha)=1 - 0.6\alpha/\alpha_{\max}
    \end{equation}
\citet{Cornwell2008} found this relations to work well with VLA images of radio galaxies. A disadvantage of this solution is that the scale bias
(\ref{eq:scale_bias}) depends on the maximum scale $\alpha_{\max}$.
This is the main reason why the original MS-CLEAN provides different
results with different scale sets. Selecting an appropriate scale
set is a tricky exercise, which is not possible to carry out in an
automated image processing. Instead of the scale bias
(\ref{eq:scale_bias}), we use a scale normalization coefficient:
\begin{equation} \label{eq:scale_norm}
    S(\alpha)=1/\sqrt{\sum\limits_{x,y}{m(\alpha)^2}}
    \end{equation}

The usage of (\ref{eq:scale_norm}) makes the scale-convolved
residuals proportional to the normalized convolution:
\begin{equation} \label{eq:convol_norm}
    I^R_\alpha S(\alpha)=\frac{I^R*m(\alpha)}{\sqrt{ \sum{m(\alpha)^2} }} \sim \frac{I^R*m(\alpha)}{\sqrt{\sum{(I^R)^2} \sum{m(\alpha)^2} }}
    \end{equation}
Equation (\ref{eq:convol_norm}) has a maximum of 1.0 when the
residual image $I^R$ exactly matches the scale-convolved PSF,
$m(\alpha)$; otherwise, it is less than 1.0. By using this
approach, the algorithm selects the scale with a scaled PSF,
$m(\alpha)$, best matching the residual image.

Thus, oppositely to the original MS-CLEAN algorithm, our scheme
adopted for SSRT data determines the position of the new clean
component by searching the maximum of the unscaled residual image,
but not by looking for the global maximum among all of the
scale-convolved residuals. We do so to follow the `up-down'
scenario, i.e., to clean the image starting from bright sources and
proceeding down to background features, which exceed the CLEAN
threshold.

Due to properties of the SSRT construction, the PSF is not
known with a required accuracy. Therefore, no modification of the
CLEAN method is able to completely reconstruct SSRT images and to
suppress all contributions from the beam's side lobes. For this
reason, a high CLEAN threshhold is applied to avoid a large number
of false sources in the `clean' map. We have empirically found for
the SSRT images the threshold of $n \, \sigma_\mathrm{HF}$, where
$\sigma_\mathrm{HF}$ characterizes high spatial-frequency
fluctuations in a raw image, and $n \approx 25$. After the cleaning
process, low-brightness features like filaments and coronal bright
points remain on the residual map being partly distorted by the side
lobes.

\subsubsection{Recovery/Filtering of Residual Map}

After the deconvolution process with the elaborated MS-CLEAN
algorithm, an additional residual filtering is performed. The
filtering should remove remaining contributions from the side lobes
and preserve all quiet solar features.

At the first step, we find areas in the image, where remaining
contributions from the side lobes are significant. To identify these
areas, we calculate a parameter characterizing typical excursions
along the directions of the residual stripes due to the contribution
from side lobes. This parameter is computed as a median value over
the standard deviations in partial regions selected by a boxcar
moving along the stripes. The areas, in which this parameter exceeds
a pre-selected threshold, are considered to be contaminated by side
lobes.

At the second step, an iterative procedure is performed, which
removes the contributions from the side lobes by using non-distorted
surrounding areas as a reference. This procedure operates in a
CLEAN-like `up-down' scenario. The CLEAN decomposes an image into a
set of `clean components' convolved with the PSF. Instead, our
iterative filter decomposes an image into a set of overlapping image
components, each of which is a smoothed sub-image multiplied by a
Hanning window. The algorithm filters out those components, which
are most likely contributions of side lobes, and reassemble the
filtered image from the remaining ones. The scheme of this iterative
procedure is as follows.

\begin{itemize}
\item Initialize
    \begin{itemize}
        \item Residual image: $I^R =$ CLEAN residual
        \item Filtered image: $I^f=0$
        \item Filter width: $\sigma_f$.
        \item Component width: $\sigma_c$.
        \item Smooth width: $\sigma_s$.
        \item High pass filtered image: $I^{hp}=I^R-I^R*G(\sigma_f)$, where $G(\sigma_f)$ is a 2d Gaussian function with a width of $\sigma_f$.
    \end{itemize}
\item Repeat
    \begin{itemize}
    \item Find the maximum of $|I^{hp}|$
    \item Extract from $I^{hp}$ a sub-image $I^{hp}_{s}$ as wide as $\sigma_w$ centered at the position of the maximum.
     \item Calculate the current image component $C_i$ by multiplying the sub-image by Hanning window and convolving it with a Gaussian function $C_i=(I^{hp}_{s}H)*G(\sigma_s)$.
     \item Subtract $C_i$ from both $I^R$ and $I^{hp}$
     \item Check if the $C_i$ position lies inside the area, where the remaining contributions from the side lobes are significant. Otherwise, add $C_i$ to the model image $I^f$.
     \item Recalculate $I^{hp}$ every 10 iterations.
    \end{itemize}
\item Until
    \begin{itemize}
     \item Either maximum iteration number reached
      \item Or $\max(|I^r|)<\mathrm{threshold}$
      \end{itemize}
\item Finalize
    \begin{itemize}
     \item Add the remaining residual $I^R$ to the filtered image $I^f$ to get the result.
      \end{itemize}

\end{itemize}

This algorithm has four important parameters. The filter width,
$\sigma_f$, sets the width of the reference area surrounding every
image component. This are should be, at least, 2--3 times wider than
the major PSF lobe (the major maximum of the interference beam
pattern). The component width, $\sigma_c$, sets the size of the
sub-images used for the decomposition. This width should slightly
exceed the main PSF lobe. The smoothing width, $\sigma_s$, is a
smoothing parameter used in the calculation of the components from
the sub-images. This width should be considerably less than the
major PSF lobe. The last parameter is the filter threshold. The
standard deviation of the CLEAN residual is a good choice for this
parameter.

\begin{figure*}
  \begin{center}
    \FigureFile(140mm,130mm){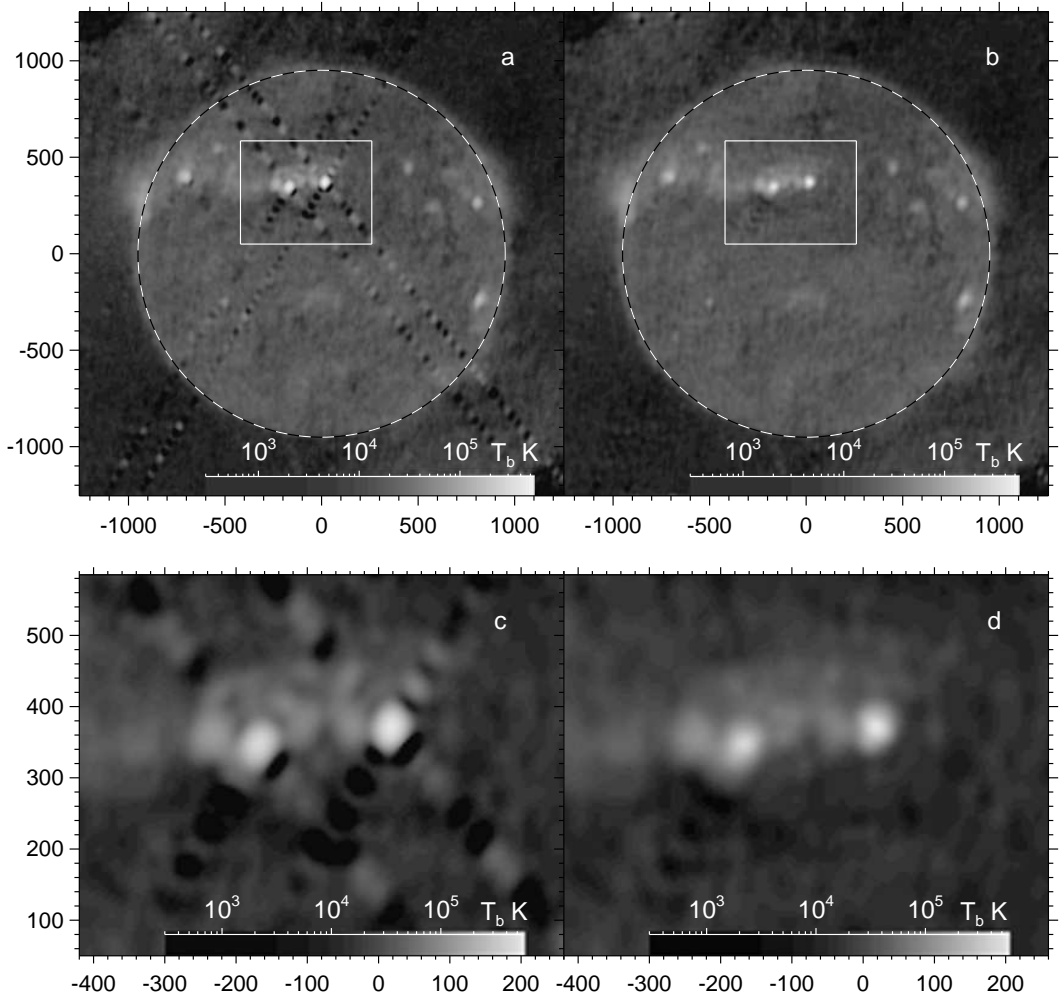}
  \end{center}
  \caption{Single SSRT images observed on 2011 May 5: dirty image (a)
and clean image (b). The lower panels show the enlarged parts of
the dirty (c) and clean (d) images marked by the white frame in
the panels above them. The axes show hereafter in similar images
the distances from the solar disk center in arc seconds. The gray
scale bars quantify the images in brightness temperatures.}
  \label{fig:clean_demo}
 \end{figure*}

\begin{figure}
  \begin{center}
    \FigureFile(75mm,57mm){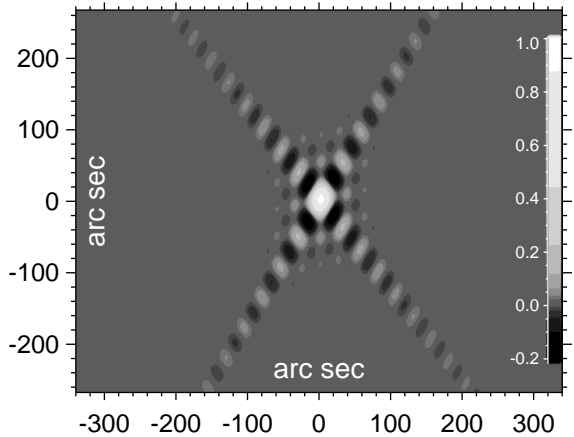}
  \end{center}
  \caption{The SSRT PSF used as an input for the deconvolution
process. The observation time and field of view correspond to those
in Figure~\ref{fig:clean_demo}c,\,d.}
  \label{fig:beam_model}
 \end{figure}

Figure~\ref{fig:clean_demo} presents an example of a dirty
single-pass SSRT image observed on 2011 May 5 (a) and the result of
its processing with deconvolution and filtering (b). The enlarged
framed area containing active region sources is shown in the lower
row (c,~d). The beam model used in the CLEAN/filtering of this image is
presented in Figure~\ref{fig:beam_model}.

To assess both qualitatively and quantitatively how realistic the
SSRT images after the CLEAN/filtering are, we compare them with NoRH
17 GHz data and extreme-ultraviolet images. To enhance the
sensitivity of the SSRT images down to $\lesssim 500$~K, we have
averaged several tens of single-pass images observed on that day
with a preparatory compensation for the solar rotation. The same
averaging procedure has been done for 46 pairs of the NoRH snapshots
produced in steps of 10~min.

\begin{figure*}
  \begin{center}
    \FigureFile(170mm,117mm){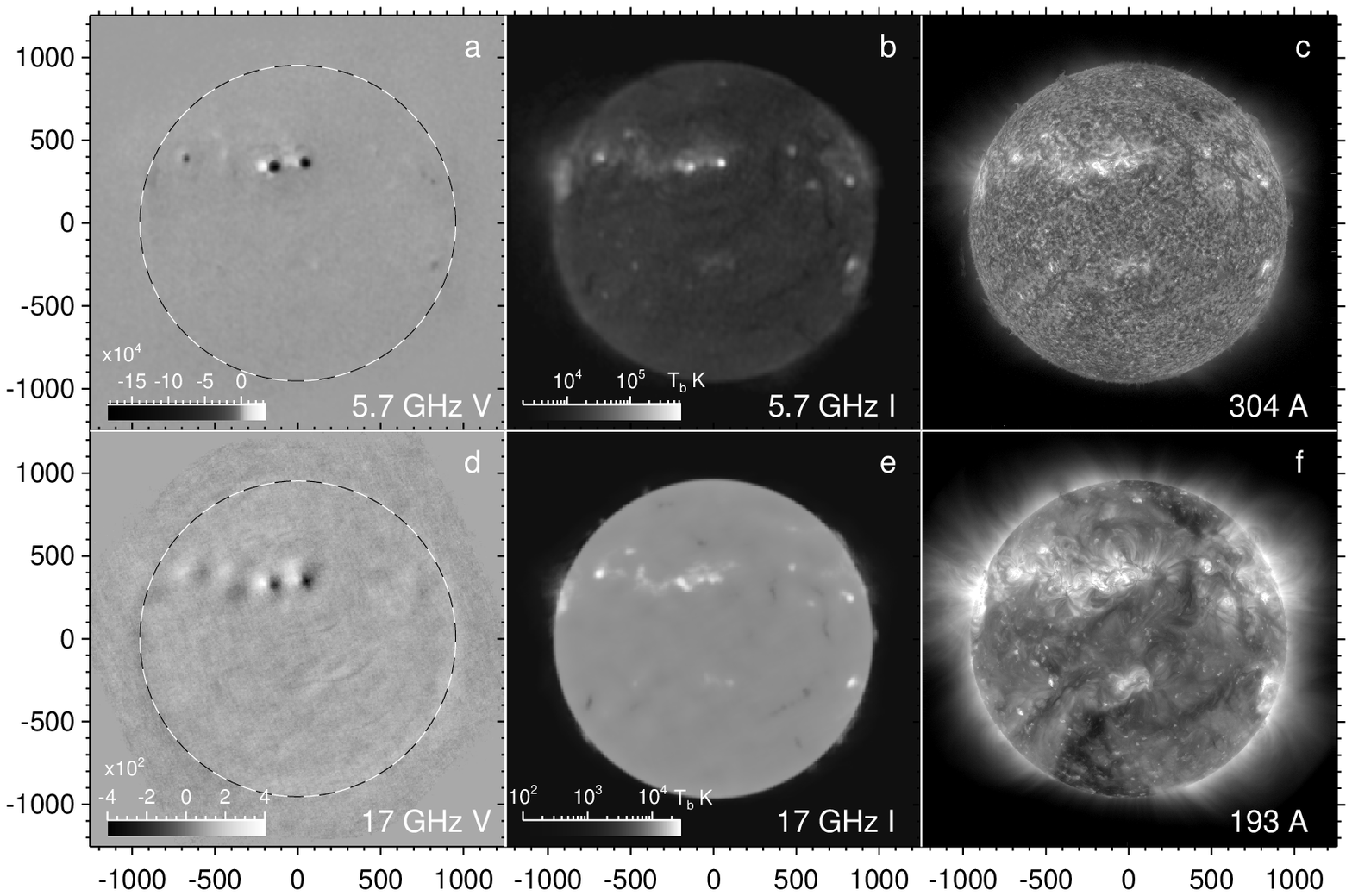}
  \end{center}
  \caption{Comparison of solar images produced on 2011 May 5 in
different spectral ranges. Top row: SSRT 5.7 GHz (a,~polarized
emission; b,~total intensity) and SDO/AIA 304~\AA\ (c). Bottom
row: NoRH 17 GHz (d,~polarized emission; e,~total intensity) and
SDO/AIA 193~\AA\ (f). All the SSRT and NoRH images shown here are
daily averages; the SDO/AIA images are snapshots. The dashed
circles in panels (a) and (d) denote the optical limb.}
  \label{fig:ssrt_norh_aia}
 \end{figure*}

We juxtapose in Figure~\ref{fig:ssrt_norh_aia} the daily-integrated
microwave images of the Sun observed on 2011 May 5 by the SSRT at
5.7 GHz (a,b) and NoRH at 17 GHz (d,e) with SDO/AIA snapshots
produced in the transition-region He\,\textsc{ii} 304~\AA\ line (c)
and in the coronal Fe\,\textsc{xii} 193~\AA\ line (f). The microwave
images present the circular polarization
(Figure~\ref{fig:ssrt_norh_aia}a,d) and total intensity
(Figure~\ref{fig:ssrt_norh_aia}b,e). Besides bright coronal
structures visible in the NoRH image, the enhanced-quality SSRT
image clearly shows chromospheric structures. Prominences and
filaments are also clearly visible.

The brightness temperature over the 17 GHz Stokes $I$ image does
not exceed 27\,000~K. The strongest polarized emission of $-366$~K
occurs in the west negatively-polarized source, in which the
degree of polarization does not exceed $-20\%$. All other sources
are polarized up to $\pm 5\%$. Thus, the polarized emission at 17
GHz in this image is most likely due to optically thin thermal
bremsstrahlung.

The brightness temperatures in the major negatively polarized
sources at 5.7~GHz reach in total intensity 0.6~MK in the East
source and 0.9~MK in the West source. Both these sources have the
degree of polarization of about 40\%. These properties indicate a
gyroresonance emission at 5.7~GHz. The dissimilarities between the
SSRT and NoRH images observed at three-times different frequencies
correspond to expectations; while the third-harmonic gyroresonance
emission occurs at 5.7~GHz in a magnetic field of 680~G, even the
fifth-harmonic emission at 17~GHz requires 1200~G at the base of
the corona.

\subsubsection{Calibration Technique}
 \label{sect:calibration}

The SSRT images are calibrated in brightness temperatures as $I =
(RCP+LCP)/2$ and $V = (RCP-LCP)/2$, respectively, referring to the
quiet Sun's brightness temperature $T_{QS} =$ 16\,000~K
\citep{Zirin1991,Borovik1994}. Initially, the technique used to
calibrate NoRH images \citep{Hanaoka1994norh} was adopted at the
SSRT. This calibration technique is based on the analysis of the
brightness distribution in an image, in which two statistical peaks
should be present. One of them corresponds to the zero sky level,
and the second one corresponds to the quiet Sun's level. The images
are calibrated by referring to the positions of the maximum values
in the two peaks.

\begin{figure}
  \begin{center}
    \FigureFile(85mm,125mm){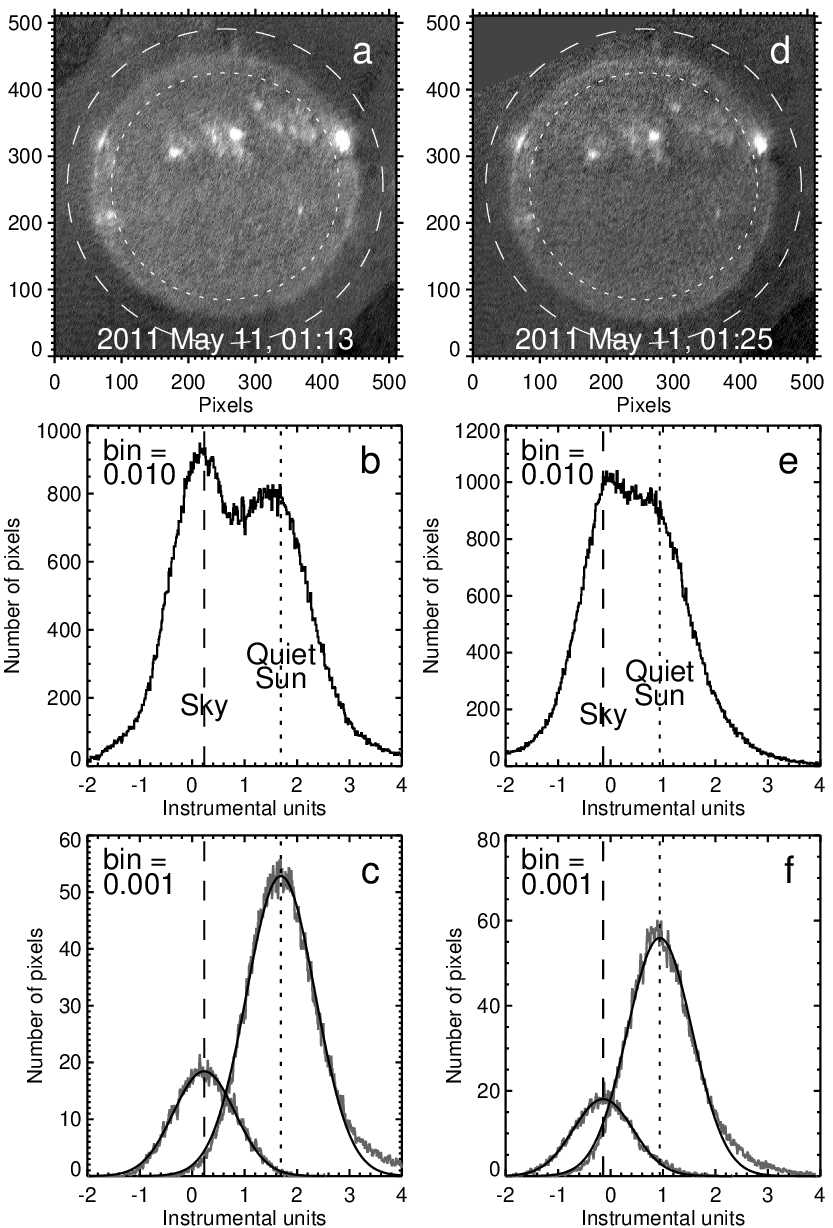}
  \end{center}
  \caption{Calibration of SSRT images illustrated with two examples
shown in the left and right columns. a),\,b):~raw images to be
calibrated. The histograms are calculated separately within the
area limited by the dotted circle (quiet Sun) and outside the
dashed circle (sky). b),\,e):~the histograms calculated over the
whole images. c),\,f):~the histograms calculated for the sky and
quiet Sun separately (gray) along with the Gaussian fit (black).}
    \label{fig:ssrt_calib_demo}
\end{figure}

Later we have elaborated this technique to improve its stability and
accuracy. They are especially critical for the SSRT images because
of their lower quality relative to NoRH ones, intrinsically broader
quiet Sun's brightness distribution at 5.7 GHz due to the larger
contribution from inhomogeneities in the chromosphere and corona,
including the higher limb brightening as discussed in
section~\ref{sect:features_in_ssrt_images}. At the present time, we
analyze the brightness distribution within the solar disk and
outside it separately, as Figure~\ref{fig:ssrt_calib_demo}
illustrates.

Two columns in the figure present two single-pass SSRT images
observed with an interval of 12 min but formed in somewhat different
conditions. The portions of the dark disks in the upper left and
lower right corners of Figure~\ref{fig:ssrt_calib_demo}a are due to
the negative response from the adjacent interference numbers. The
blank gray triangle area in the upper left corner of
Figure~\ref{fig:ssrt_calib_demo}d is due to missing data. The middle
panels show the histograms computed over the whole images. The lower
panels show the gray histograms computed separately for the
quiet-Sun area inside the dotted circle shown in the upper panels
and the sky area computed outside the dashed circle. Each of the
peaks of separate histograms shown in the lower panels is fitted
with a Gaussian (black curves), and their calculated positions are
used to calibrate the images.

The position of the quiet Sun's level is displaced with respect to
the apparent maximum of the right peak in
Figure~\ref{fig:ssrt_calib_demo}b, while the two peaks in the
histogram in Figure~\ref{fig:ssrt_calib_demo}e is difficult to
recognize. These examples illustrate the robustness of the advanced
technique.

This calibration technique has turned out to be useful for NoRH
images also. Their calibration stability becomes problematic, when
the brightness temperatures of sources of interest are comparable
with the clean threshold. Our elaborated technique allows us to
improve the calibration quality of NoRH images considerably.

\begin{figure}
  \begin{center}
    \FigureFile(57mm,125mm){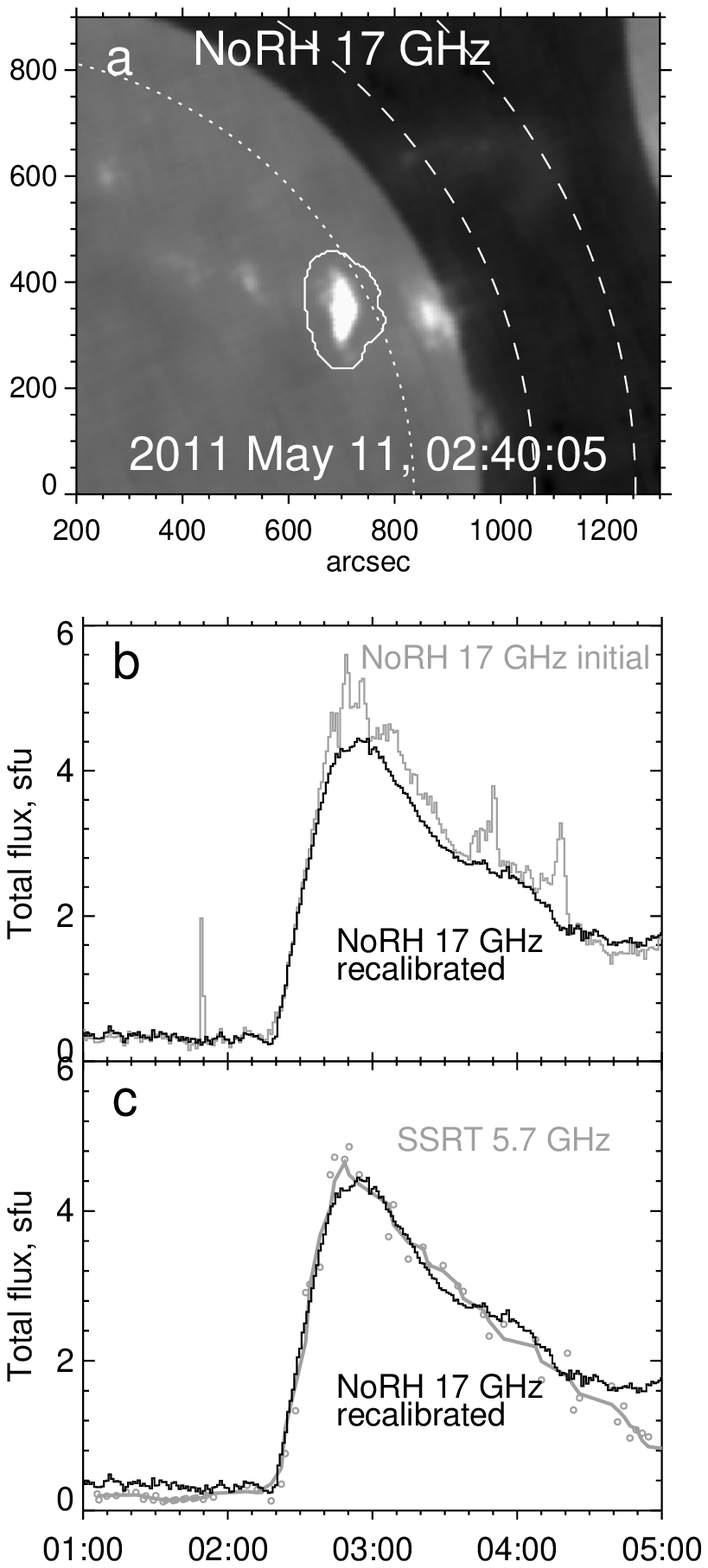}
  \end{center}
  \caption{Recalibration of NoRH 17 GHz data with the technique
presented in Figure~\ref{fig:ssrt_calib_demo}. a)~A portion of a
NoRH image with the contoured area of the thermal microwave source.
The histograms were calculated separately within the dotted circle
(quiet Sun) and within the wide ring limited by the dashed circles
(sky). b)~Total flux time profiles computed within the contour from
the initial NoRH images (gray) and after recalibration (black).
c)~Comparison of the total flux time profiles computed from NoRH
images (black) and SSRT images (gray circles; solid gray curve is
smoothed over 5 points).}
    \label{fig:norh_recalib_comp}
\end{figure}

Figure~\ref{fig:norh_recalib_comp} illustrates the result of
recalibration of the NoRH images produced during a very weak burst
caused by a non-active-region filament eruption on 2011 May 11. The
histogram for the quiet Sun was computed within the dotted circle,
and the histogram for the sky was computed within the wide ring
between the dashed circles. The gray time profile in
Figure~\ref{fig:norh_recalib_comp}b was computed over the burst
source area inside the solid contour from a set of images with 10~s
integration time produced by the Koshix program in steps of 10~s.
The black time profile was computed from recalibrated images. The
total flux of the Sun at 17 GHz on that day was 630 sfu.
Figure~\ref{fig:norh_recalib_comp}c compares the time profiles of
the burst recorded with NoRH at 17 GHz and SSRT at 5.7 GHz. Their
close correspondence to each other certifies almost purely thermal
nature of the microwave burst. On the other hand, this example
demonstrates that measurements from NoRH data can be still improved.

The elaborated techniques described in the present
Section~\ref{sect:imaging_techniques} have automatically produced
enhanced-quality SSRT images for many days of observations. Several
observational intervals still remain, for which attempts to produce
such images were not successful; routine SSRT data can be used for
these days. The new data set processed to the present time offers
wide opportunities to study low-brightness features at 5.7~GHz. Some
examples are considered in
section~\ref{sect:features_in_ssrt_images}. The data base of
enhanced-quality daily averaged SSRT images is regularly updated and
available at http:/\negthinspace/ssw.iszf.irk.ru.

\section{Computation of Synoptic Maps from SSRT Images}
 \label{sect:synoptic}

Daily radio brightness distributions of total intensity and
polarized emission accumulated in our data base supplemented with
synoptic distributions for the Carrington rotations of the Sun. In
computation of a required rotation we use a set of full-disk daily
images, which correspond to a time interval starting from the
middle of the preceding rotation and ending at the middle of the
next one. In the computation of such a map, we use a synoptic grid
of $[1080 \times 540]$ samples on the surface of the Carrington
longitude and the sine of the latitude within the intervals of
$[1/6,\ 360-1/6]$ and $[-1+1/540,\ 1-1/540]$, respectively. Each
node $(l, j)$ of the grid is associated with a Julian day
$JD_\mathrm{net}^{\mathrm{SYN}}$ corresponding to the Carrington
longitude of the node. To calculate the necessary synoptic
quantity in a selected node $(l, j)$ of the synoptic map, the
following operations are carried out:

\begin{enumerate}

\item
 From each full-disk data array, a value is interpolated to each current node
 of the synoptic grid, if this node belongs to the current plane of
 the sky. As a result, two one-dimensional arrays are formed. One
 array contains interpolated values $f$, and another array
 contains Julian dates $JD$ corresponding to the observation times of the
 full-disk images, from which the values to be interpolated were
 selected. Both one-dimensional arrays have the same number of
 elements $M$, and the index $s = 1,2, ... M$.
 Generally, $M < N$, where $N$ is the total number of
 the full-disk source images, and $M$ depends on the position of
 the node $(l, j)$.

\item
 The ultimate value of a synoptic parameter for a node $(l, j)$ is
 calculated as a weighted average:

\begin{equation}
 \label{eq:fig}
F^{\mathrm{SYN}}\left( i,j\right)
=\frac{\sum\limits_{s=1}^{M}f\left( s\right) a\left( s\right)
}{\sum\limits_{s=1}^{M}a\left( s\right) },
\end{equation}
where $a\left( s\right) =\exp \left\{ -\left[ 12.9\,
\frac{JD_\mathrm{net}^{\mathrm{SYN}}\left( i,j\right) -JD\left(
s\right) }{27.275}\right] ^{2}\right\} $; 12.9 is a weighting factor
optimized experimentally.

\end{enumerate}

Equation (\ref{eq:fig}) provides the largest contribution from
those pixels in a full-disk image, whose distance is minimum from
the central meridian corresponding to the Carrington longitude of
the node $(l, j)$. The calculated synoptic data for each
Carrington rotation are recorded into FITS files and posted to the
main data base.

All microwave images are combined with a corresponding potential
magnetic field. The graphic files are produced for both daily
full-disk images and synoptic maps and posted to the same data base.
The magnetic field is calculated from the coefficients of the
spherical harmonic decomposition contained in the BD-Monitoring data
base (http:/\negthinspace/bdm.iszf.irk.ru/).

\section{Low-brightness Features in SSRT Images}
 \label{sect:features_in_ssrt_images}

Microwave observations of the quiet Sun have been reviewed by
\citet{Kundu1982} and more recently by \citet{Shibasaki2011}.
Active low-brightness features such as eruptive prominences and
filaments in microwave SSRT and NoRH images have been also
discussed previously (e.g.,
\cite{Hanaoka1994fil,Uralov2002,Shimojo2006};
\authorcite{Grechnev2006erupt}
\yearcite{Grechnev2003,Grechnev2006quiet,Grechnev2006erupt}; and
others). Here we briefly present low-brightness features in
enhanced-quality SSRT images, which can contribute to further
investigations into these issues.

\subsection{Brightness Temperature Radial Distribution}
 \label{sect:radial_distribution}

\begin{figure}
  \begin{center}
    \FigureFile(85mm,85mm){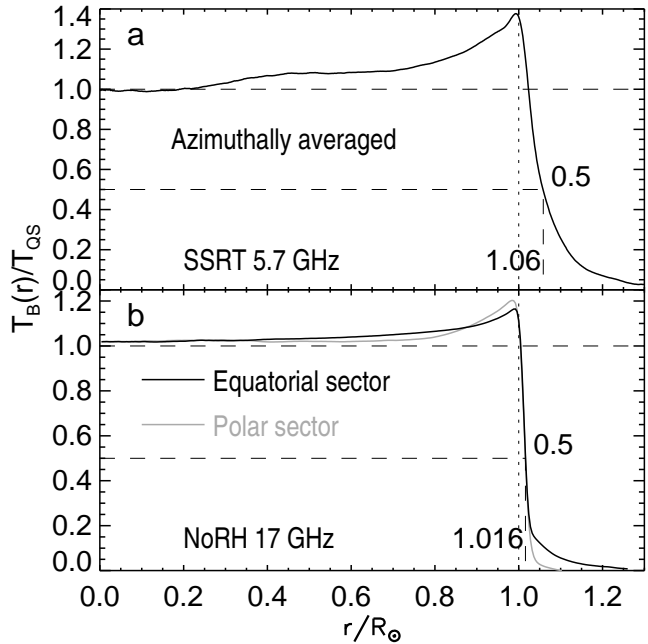}
  \end{center}
  \caption{Averaged radial distribution of the brightness
temperature: a)~SSRT, 5.7~GHz, azimuthally averaged distribution;
b)~NoRH, 17~GHz, in the equatorial (black) and polar (gray)
sectors of $15^{\circ}$ widths.}
  \label{fig:quiet_sun}
 \end{figure}

The average brightness temperature distribution was measured from
296 images produced by the SSRT from 2010 April 1 to 2010 September
30 and from 2011 April 1 to 2011 July 30. Before the measurements,
daily averaged SSRT images were additionally averaged over eight
days. Then the averaged images were analyzed with ring scanning. The
ring had a width of one pixel and progressively increasing radius.
Data within each ring were analyzed statistically. All pixels with
values deviating from an average value by more than $\pm 2\sigma$
were discarded, and a new average calculated over remaining pixels
was assigned as a brightness temperature at the current radius. This
technique does not take account of the elliptical shape of the radio
brightness distribution, and the result depends on the current
situation on the Sun. The result is shown in
Figure~\ref{fig:quiet_sun}a. This is the first attempt to measure
this distribution from two-dimensional SSRT images and can be
elaborated in future.

For comparison we also computed the average brightness temperature
distribution from NoRH 17 GHz snapshots (one per day).
Measurements from NoRH data covered the periods from 1992 July to
1995 August and from 1995 November to 1996 April (totally 44
months, about 1320 images). We carried out the measurements from
monthly averaged NoRH images. In this case, the higher data
quality and uniformity allowed us to analyze the brightness
temperature distribution with a more complex algorithm in 24
sectors separately, each of $15^{\circ}$. The results are shown in
Figure~\ref{fig:quiet_sun}b.

The central part of the distribution at 5.7 GHz corresponds to the
`bald' Sun, while at $r > 0.2R_{\odot}$ the brightness temperature
increases by about 8\% probably due to the contribution from active
regions at mid latitudes. This contribution is present up to the
limb because of the azimuthal averaging. The position of the
inflection probably depends on the phase of the solar activity. The
radio radius at a 0.5 level of the brightness at the disk center is
about $\approx 1.06R_{\odot}$. The limb brightening spike reaches
$1.38 T_{\mathrm{QS\ 5.7}} \approx 22\,000$~K. These parameters are
close to the measurements by \citet{Kundu1979} with WSRT at 6~cm
that gave the radio radius of $\approx 1.05R_{\odot}$ and the limb
brightening up to 1.4--1.5 of the brightness temperature near the
solar disk center. Our results are also mostly close to those
measured by \citet{Krissinel2005} from one-dimensional SSRT scans
observed in the solar minimum. Unlike the results obtained from the
one-dimensional SSRT data, a considerable polar limb brightening
seems to be clearly present in most two-dimensional SSRT images, in
accordance with the measurements at 6 cm reported by
\citet{Kundu1979}. This difference might be due to the solar cycle
variations found by
\authorcite{Selhorst2004} (\yearcite{Selhorst2004,Selhorst2011}) for
the quiet Sun at 17 GHz.

Our measurements show that the radio radius at 17 GHz amounts to
$1.016R_{\odot}$ (Figure~\ref{fig:quiet_sun}b). The limb spike
reaches $1.16 T_{\mathrm{QS\ 17}}$ in the equatorial sector and
$1.20 T_{\mathrm{QS\ 17}}$ in the polar sector ($T_{\mathrm{QS\ 17}}
= 10\,000$~K). These parameters of the limb brightening are
consistent with those shown by \citet{Nindos1999} and the
statistical results of \authorcite{Selhorst2004}
(\yearcite{Selhorst2004,Selhorst2011}). All of these quantities were
measured directly from SSRT and NoRH images without any correction
for deconvolution.

\subsection{Supergranulation Network}
 \label{sect:network}

\begin{figure*}
  \begin{center}
    \FigureFile(170mm,130mm){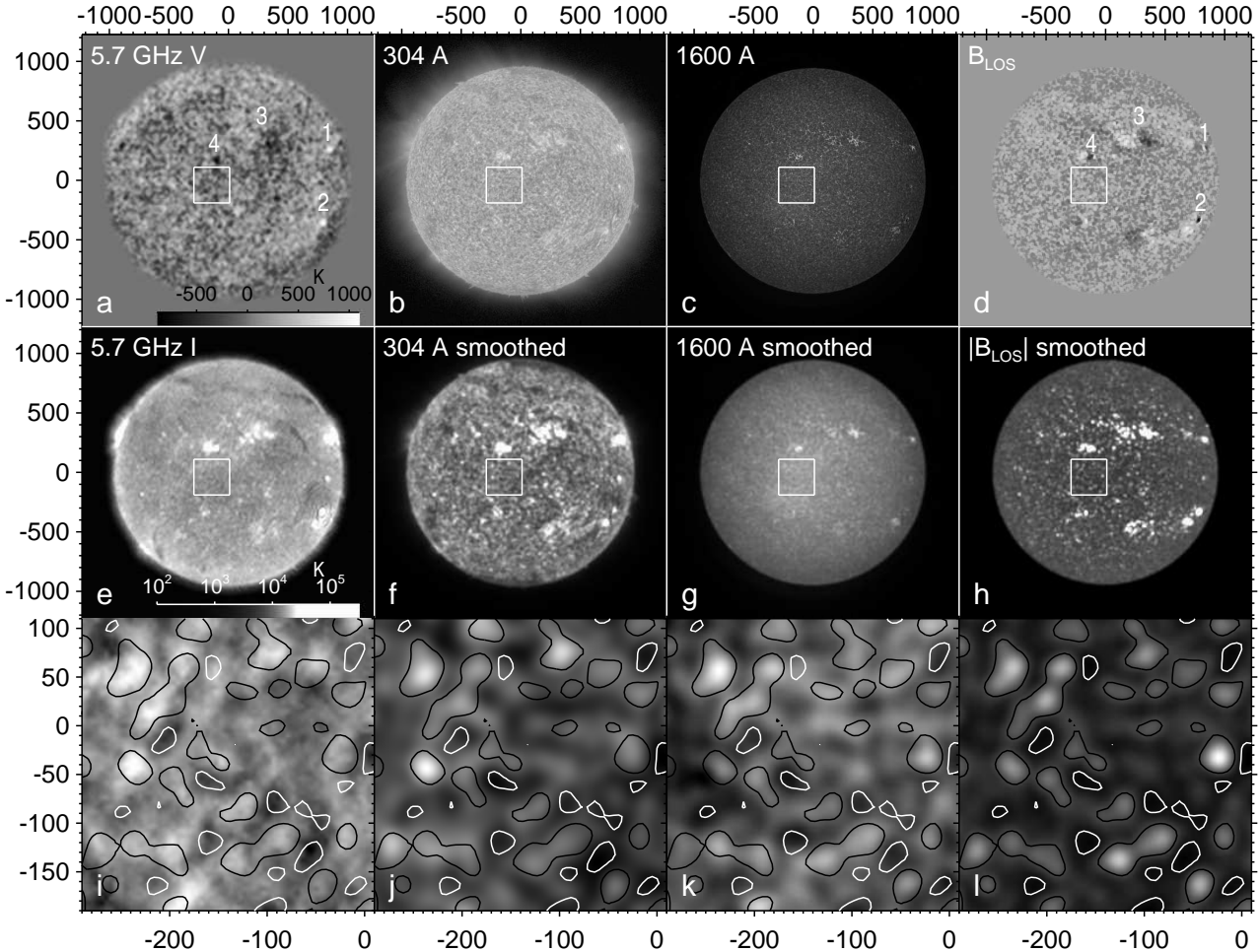}
  \end{center}
  \caption{Supergranulation network in microwave and EUV images (SDO/AIA)
and the magnetogram (SDO/HMI) observed on 2010 July 7 (all
displayed strongly nonlinearly). All the images in each column
present the same wavelength. The images in the middle row present
the images from the upper row with a spatial resolution reduced to
the SSRT one. The lower row shows an enlarged area of the images
in the middle row denoted by the white frame. The contours
overlaid on all the images in the lower row correspond to the
levels of 4.5 and 6.8~G in the smoothed $|B_\mathrm{LOS}|$
magnetogram.}
  \label{fig:network}
 \end{figure*}

It is possible to study the network in microwaves from SSRT images,
although with a poorer spatial resolution of order $25^{\prime
\prime}$. The relative stability of the network on the time scale
of several hours allows one to use daily averaged SSRT images to
improve the sensitivity. Figure~\ref{fig:network} presents a
comparison of the network appearance in the SSRT image averaged
over 5 hours on 2010 July 7 along with the snapshots produced at
about 05:00 by SDO/AIA in 304~\AA, 1600~\AA, and the SDO/HMI
magnetogram. We have additionally considered an averaged NoRH 17
GHz image and an SDO/AIA 171~\AA\ snapshot. We do not show them
for the lack of any manifestations of the network in these images.

The upper row of Figure~\ref{fig:network} shows the SSRT Stokes
$V$ image (a), raw SDO/AIA 304~\AA\ (b) and 1600~\AA\ (c) images,
and the line-of-sight SDO/HMI magnetogram ($B_\mathrm{LOS}$, d).
To reveal weak polarized emission in the SSRT image, we have
smoothed it with a boxcar of $7 \times 7$ pixels size and limited
the brightness from above by 1\,100~K. Actually source 1 reaches
3\,000~K, source 2 reaches 15\,400~K. The signs of the
polarization of sources 1 and 2 are reverted and correspond to the
o-mode emission, which is a well-known effect typically observed
at low microwaves close to the limb. An extended plage region 3
with brightness temperatures exceeding $\pm 1000$~K and a compact
source 4 ($-1500$~K) are polarized in the sense of the x-mode
emission being located far from the limb.

All the images in the middle row, Figure~\ref{fig:network}e--h,
are presented in the form corresponding to the SSRT resolution.
These are the SSRT Stokes $I$ image (e), SDO/AIA 304~\AA\ (f) and
1600~\AA\ (g) images smoothed by convolving with the SSRT beam
pattern, and the absolute value of the line-of-sight SDO/HMI
magnetogram ($|B_\mathrm{LOS}|$, h).

The images in the lower row, Figure~\ref{fig:network}i--l, are
enlarged portions of the images just above denoted by the frames:
Stokes $I$ at 5.7 GHz (i), SDO/AIA 304~\AA\ (j), 1600~\AA\ (k),
and $|B_\mathrm{LOS}|$ (l). The contours on top of all of the
images are the same and correspond to the levels of 4.5~G (white)
and 6.8~G (black) in the smoothed $|B_\mathrm{LOS}|$ magnetogram.
These contours facilitate visual comparison of the images with the
$|B_\mathrm{LOS}|$ magnetogram.

The SSRT Stokes $I$ image presents a range of brightness
temperatures from 11\,000~K to 18\,500~K. Comparison of dark and
bright features within the contours in the SSRT image with other
images shows a good correspondence between some features (e.g., in
the upper left corner) and a poor correspondence for some others
(e.g., the brightest feature in the right middle part of the
magnetogram). Nevertheless, an overall structural correspondence
appears to be present. A careful examination can reveal extra
matching features, which are not outlined with the contours. A
typical brightness temperature of dark network elements at 5.7 GHz
is about 12\,500~K, while that of bright elements is about
18\,000~K. Thus, the SSRT image shows a contrast of the network of
about 1.44:1. This is somewhat less than the contrast estimated by
\citet{Kundu1979} at 6~cm (1.7:1); the difference can be due to a
$\sim30$\% larger opacity to the free-free emission at 6~cm (WSRT) with
respect to 5.2~cm (SSRT) as well as our usage of direct
measurements without any correction for deconvolution. The cell
sizes correspond to those observed by \citet{Kundu1979} at 6~cm,
$\sim (40-50)^{\prime \prime}$.

To assess the correspondence between the SSRT and SDO/AIA images quantitatively, we
have calculated the Pearson correlation coefficients for pairs of the images within the quite Sun area at the solar disk centre. The calculation was done  for 10 image samples obtained at June-July 2010. The EUV images and $|B_\mathrm{LOS}|$ magnetograms were smoothed and resized to match spatial resolution of the SSRT. The results are summarised in Table~\ref{tab:Pearson}.

\begin{table*}
    \centering
        \begin{tabular}{c|c|c|c|c|c}
                &SSRT & HMI &AIA 1600   \AA &AIA 304    \AA &AIA 171\AA \\
                \hline
SSRT    &   ---                             &   $0.59 \pm 0.07$ &   $0.47 \pm 0.05$     &$0.68 \pm 0.08$        &   $0.50  \pm 0.15$ \\
HMI     & $0.59 \pm 0.07$   &   ---                         & $0.61 \pm 0.05 $  &$  0.73 \pm 0.03$  &   $0.42 \pm 0.10$ \\

        \end{tabular}
    \caption{ Pearson correlation coefficients  for SSRT, SDO/AIA images, and SDO/HMI magnetograms.}
    \label{tab:Pearson}
\end{table*}

The low correlation for the 171~\AA\ coronal channel is expected: while the
magnetic field is stronger at the footpoints of coronal loops, their
brightness is maximum between the footpoints, as \citet{Benz1997}
proposed. A noticeable coronal contribution at 5.7~GHz can be also
responsible for the decreased correlation of the network shown by
the SSRT with the magnetogram. The Pearson correlation coefficient
between the SSRT image and the 304~\AA\ image is the largest one,
0.68, and considerably less between the SSRT image and the 171~\AA\
image, 0.5, but still higher than between the 171~\AA\ image and
the magnetogram. Note that bright coronal features look generally
similar at 5.7 GHz and in 171~\AA\ as well as 195~\AA\ images
(\authorcite{Grechnev2003}
\yearcite{Grechnev2003,Grechnev2006quiet}).

These facts confirm that
SSRT images at 5.7~GHz show a superposition of chromospheric and
coronal contributions supplemented with compact gyromagnetic
sources. Note also that the difference between the bright and dark
network elements in SSRT images of 5500~K is well above the SSRT
sensitivity (1500~K in a single image), which allows one to analyze
the network manifestations on considerably shorter time scales than
5~hours used here.

We also analyzed a NoRH 17~GHz image averaged over the whole day
of observations, but the supergranulation network was not
detectable in this image for an obvious reason. The opacity of
microwave sources at 17~GHz is less than at 5.7~GHz by about one
order of magnitude, so that the range of brightness temperatures
of 8\,000~K in the SSRT image in Figure~\ref{fig:network}i is
expected to be at 17~GHz well below the CLEAN threshold (3000~K;
it cannot be reduced considerably).

Finally we note that the quite Sun's features discussed in this
section determine an intrinsically wider histogram of the
brightness temperature distribution at 5.7~GHz with respect to
17~GHz. The FWHM width of the histogram computed for the small
region in Figure~\ref{fig:network}i is about 2\,500~K. The
center-to-limb brightness temperature variation still increases
the width. These circumstances offer additional requirements to
the calibration technique at 5.7~GHz, as mentioned in
section~\ref{sect:calibration}.

The beacon results presented here promise a considerably larger
scientific outcome of the observations made with the SSRT.
Elaboration of these results can contribute to further
investigations into important issues of the solar physics.

\section{Summary and Concluding Remarks}
 \label{sect:summary}

Several dozens of single-pass solar images have been produced by the
SSRT every day. The elaborated techniques presented here have
allowed us to deconvolve the images with restoration of
low-brightness features in a fully automatic regime. Visual and
quantitative comparisons with data from different spectral ranges
and with the reported results previous studies at close wavelengths
confirm reliability and high quality of the data produced with our
new techniques. So far, the state-of-art does not allow us to
produce high-quality data from all of the single-pass images. We
hope to overcome this limitation.

A non-interrupted sequence is available of one single-pass SSRT
image per day observed from 2000 May up to the most recent day.
These images are produced with the routine SSRT CLEAN technique,
which provides images of active regions, but does not reveal
low-brightness features below the CLEAN threshold.

The daily averaged images with the enhanced dynamic range of $\sim
30$ dB and brightness sensitivity $\lesssim 500$~K have been
produced for more than 80 months of SSRT observations from 1999
August to 2013 February (years 2007 and 2008 have not yet been
successfully processed). These images allow one to study
quasi-stationary solar features at 5.7~GHz such as the radial
brightness temperature distribution, supergranulation network,
coronal holes, bright points, filaments and prominences, and others.
These data are easily accessible via the Internet.

The enhanced-quality single-pass SSRT images are also promising for
studies of eruptive prominences and filaments. Depending on a
particular situation, they can be observed in SSRT images up to
distances exceeding $2R_{\odot}$ from the solar disk center, i.e.,
with an overlap with the LASCO/C2 field of view. The list of
eruptive prominences and filaments observed with the SSRT contains
89 events. Very few of these events have been analyzed so far.

The initial operating principle of the SSRT (see
Section~\ref{sect:introduction}) has allowed us to produce images of
rather high quality, but restricted the opportunities in the
following aspects. The single-frequency capability of the SSRT is
the first limitation. Second, due to the direct imaging, the
receiver system is facing the whole dynamic range of the brightness
temperatures that can considerably exceed $10^{3}$ during flares.
For this reason, it is very difficult to observe flares with the
SSRT. And, third, each image of the Sun was produced by using the
Earth's rotation taking 3--5~min, so that the snapshot mode was
impossible. An obvious way to overcome the second and third
limitations is the change of the operating principle to the parallel
synthesis. The implementation of the multi-frequency capability has
also been developed.

In 2013 July the SSRT has terminated observations based on the
initial operating principle. The SSRT is currently under the upgrade
to a multi-frequency (4--8~GHz) radio heliograph. The new-generation
solar radio telescope would consist of 96 antennas and occupy the
East--West--South sub-array of the SSRT. A 10-antenna prototype of
the radio heliograph was developed \citep{Lesovoi2012}. The
following design questions have been resolved with the prototype: a
suitable dual polarization feed; transmission of RF signal from the
antennas to a working building; and, finally, interconnection of
digital receivers and the correlator. The expected brightness
temperature sensitivity of the radio heliograph is about 100~K in
the snapshot image. The expected spatial resolution would be up to
$13^{\prime \prime}$. The accuracy in the measurements of the degree
of polarization is assumed to be up to a few percents.

\bigskip

Acknowledgements. We thank S.~V.~Lesovoi, A.~M.~Uralov,
A.~T.~Altyntsev, B.~B.~Krissinel, K.~Shibasaki,
C.~E.~Alissandrakis, and S.~M.~White for fruitful discussions and
assistance. We are indebted to the referee, Prof. Kameno, for
useful remarks and encouraging suggestions. We thank colleagues
from Nobeyama Solar Radio Observatory (NAOJ) operating NoRH and
NoRP, and colleagues operating the Siberian Solar Radio Telescope.
We are grateful to the SDO/AIA and SDO/HMI consortia for their
open data policy.

This study was supported by the Russian Foundation of Basic
Research under grants 12-02-00037, 12-02-33110-mol-a-ved,
12-02-31746-mol-a, and 13-02-90472-ukr-f-a; the Program of base
fundamental research of the SB RAS  No. 16.1, and the Russian
Ministry of Education under projects 8407 and 14.518.11.7047.



\begin{thebibliography}{}

\bibitem[Alissandrakis et al.(1992)]{Alissan1992} Alissandrakis,
C.~E., Lubyshev, B.~I., Smol'kov, G.~I., et al.\ 1992, \solphys,
142, 341

\bibitem[Altyntsev et al.(1994)]{Altyntsev1994}
Altyntsev, A. T., Grechnev, V. V., Kachev, L. E., Lesovoi, S. V.,
Mansyrev, M. I., Molodyakov, S. A., Platonov, A. V., Saenko, I.
I., Smolkov, G. Y., Sych, R. A., Treskov, T. A., Zandanov, V. G.,
Esepkina, N. A.\ 1994, \aap, 287, 256

\bibitem[Benz et al.(1997)]{Benz1997}
Benz, A. O., Krucker, S., Acton, L. W., Bastian T. S.\ 1997, \aap,
320, 993

\bibitem[Borovik(1994)]{Borovik1994}
Borovik, V. N.\ 1994, Lecture Notes in Physics, Advances in Solar
Physics. Catania, Italy, 11--15 May, 1993, 432, 185

\bibitem[Bogod \& Korolkov(1975)]{Bogod1975}
Bogod, V.~M., \& Korolkov, D.~V.\ 1975, Pisma v Astronomicheskii
Zhurnal, 1, 25

\bibitem[Bogod(1978)]{Bogod1978}
Bogod, V.~M.\ 1978, Soobshcheniya Spetsial'noj Astrofizicheskoj
Observatorii, 23, 22

\bibitem[Cornwell(2008)]{Cornwell2008}
Cornwell, T. J.\ 2008, IEEE Journal of Selected Topics in Signal
Processing, 2, 793


\bibitem[Enome(1995)]{Enome1995}
Enome, S.\ 1995, Coronal Magnetic Energy Releases, 444, 35

\bibitem[Grechnev et al.(2003)]{Grechnev2003}
Grechnev, V. V., Lesovoi, S. V., Smolkov, G. Y., Krissinel, B. B.,
Zandanov, V. G., Altyntsev, A. T., Kardapolova, N. N., Sergeev, R.
Y., Uralov, A. M., Maksimov, V. P., Lubyshev, B. I.\ 2003,
\solphys, 216, 239

\bibitem[Grechnev et al.(2006a)]{Grechnev2006quiet}
Grechnev, V. V., Uralov, A. M., Maksimov, V. P., Zandanov, V. G.,
Smolkov, G. Y., Altyntsev, A. T., Krissinel, B. B., Kardapolova, N.
N., Lesovoi, A. V., Lubyshev, B. I., Prosovetsky, D. V., Rudenko, G.
V.: 2006a, Solar Physics with the Nobeyama Radioheliograph, 101

\bibitem[Grechnev et al.(2006b)]{Grechnev2006erupt}
Grechnev, V. V., Uralov, A. M., Zandanov, V. G., Baranov, N. Y.,
Shibasaki, K.\ 2006b, \pasj, 58, 69

\bibitem[Hanaoka et al.(1994a)]{Hanaoka1994norh}
Hanaoka, Y., Shibasaki, K., Nishio, M., Enome, S., Nakajima, H.,
Takano, T., Torii, C., Sekiguchi, H., Bushimata, T., Kawashima,
S., Shinohara, N., Irimajiri, Y., Koshiishi, H., Kosugi, T.,
Shiomi, Y., Sawa, M., Kai, K.\ 1994a, Proceedings of Kofu
Symposium, 35

\bibitem[Hanaoka et al.(1994b)]{Hanaoka1994fil}
Hanaoka, Y., Kurokawa, H., Enome, S., Nakajima, H., Shibasaki, K.,
Nishio, M., Takano, T., Torii, C., Sekiguchi, H., Kawashima, S.,
Bushimata, T., Shinohara, N., Irimajiri, Y., Koshiishi, H.,
Shiomi, Y., Nakai, Y., Funakoshi, Y., Kitai, R., Ishiura, K.,
Kimura, G.\ 1994b, \pasj, 46, 205

\bibitem[H{\"o}gbom(1974)]{Hogbom1974}
H{\"o}gbom, J. A.\ 1974, \aaps, 15, 417

\bibitem[Kochanov et al.(2011)]{Kochanov2011}
Kochanov, A. A., Anfinogentov, S. A., Prosovetsky D. V.\ in Solar
and Solar-Terrestrial Physics 2011, Proc. 16-th Pulkovo
International Conference on Solar Physics (GAO RAN, St.
Petersburg, Pulkovo, 2011), 227

\bibitem[Koshiishi et al.(1994)]{Koshiishi1994}
Koshiishi, H., Enome, S., Nakajima, H., et al.\ 1994, \pasj, 46, L3

\bibitem[Krissinel'(2005)]{Krissinel2005}
Krissinel', B. B.\ 2005, Astron. Rep., 49, 939

\bibitem[Krissinel et al.(2000)]{Krissinel2000}
Krissinel, B. B., Kuznetsova, S. M., Maksimov, V. P., Prosovetsky,
D. V., Grechnev, V. V., Stepanov, A. P., Shishko, L. F.\ 2000,
\pasj, 52, 909

\bibitem[Kundu(1982)]{Kundu1982}
Kundu, M. R.\ 1982, Reports on Progress in Physics, 45, 1435

\bibitem[Kundu et al.(1979)]{Kundu1979}
Kundu, M. R., Rao, A. P., Erskine, F. T., Bregman, J. D.\ 1979,
\apj, 234, 1122

\bibitem[Lesovoy(2002)]{Lesovoy2002}
Lesovoy, S. V.\ 2002, Radiophysics and Quantum Electronics, 45, 865

\bibitem[Lesovoi et al.(2012)]{Lesovoi2012}
Lesovoi, S. V., Altyntsev, A. T., Ivanov, E. F., Gubin, A.V.\ 2012,
\solphys, 280, 651

\bibitem[Maksimov et al.(2006)]{Maksimov2006}
Maksimov, V. P., Prosovetsky, D. V., Grechnev, V. V., Krissinel,
B. B., Shibasaki, K.\ 2006, \pasj, 58, 1

\bibitem[Nakajima et al.(1994)]{Nakajima1994}
Nakajima, H., Nishio, M., Enome, S., Shibasaki, K., Takano, T.,
Hanaoka, Y., Torii, C., Sekiguchi, H. et al.\ 1994, Proc. IEEE,
82, 705

\bibitem[Nakajima et al.(1985)]{Nakajima1985}
Nakajima, H., Sekiguchi, H., Sawa, M., Kai, K., Kawashima, S.\
1985, \pasj, 37, 163

\bibitem[Nindos et al.(1999)]{Nindos1999}
Nindos, A., Kundu, M. R., White, S. M., Gary, D. E., Shibasaki, K.,
Dere, K. P.\ 1999: \apj, 527, 415


\bibitem[Rudenko(2001)]{Rudenko2001}
Rudenko, G. V.\ 2001, Solar Phys., 198, 5

\bibitem[Selhorst et al.(2004)]{Selhorst2004}
Selhorst, C.~L., Silva, A.~V.~R., Costa, J.~E.~R.\ 2004, \aap,
420, 1117

\bibitem[Selhorst et al.(2011)]{Selhorst2011} Selhorst, C.~L.,
Gim{\'e}nez de Castro, C.~G., V{\'a}lio, A., Costa, J.~E.~R.,
Shibasaki, K.\ 2011, \apj, 734, 64

\bibitem[Shibasaki et al.(2011)]{Shibasaki2011}
Shibasaki, K., Alissandrakis, C. E., Pohjolainen, S.\ 2011,
\solphys, 273, 309

\bibitem[Shimojo et al.(2006)]{Shimojo2006}
Shimojo, M., Yokoyama, T., Asai, A., Nakajima, H., Shibasaki, K.\
2006, \pasj, 58, 85

\bibitem[Smolkov et al.(1986)]{Smolkov1986}
Smolkov, G. I., Pistolkors, A. A., Treskov, T. A., Krissinel, B.
B., Putilov, V. A.\ 1986,  \apss, 119, 1

\bibitem[Steer et al.(1984)]{Steer1984}
Steer, D. G., Dewdney, P. E., Ito, M. R.\ 1984, \aap, 137, 159

\bibitem[Uralov et al.(1998)]{Uralov1998}
Uralov, A. M., Grechnev, V. V., Lesovoi, S. V., Sych, R. A.,
Kardapolova, N. N., Smolkov, G. Ya., Treskov, T. A.\ 1998,
\solphys, 178, 557

\bibitem[Uralov et al.(2002)]{Uralov2002}
Uralov, A. M., Lesovoi, S. V., Zandanov, V. G., Grechnev, V. V.\
2002, \solphys, 208, 69


\bibitem[Uralov et al.(2008)]{Uralov2008}Uralov, A. M.,
Grechnev, V. V., Rudenko, G. V., Rudenko, I. G., Nakajima, H.\
2008, \solphys, 249, 315


\bibitem[Zirin et al.(1991)]{Zirin1991}
Zirin, H., Baumert, B. M., Hurford, G. J.\ 1991, \apj, 370, 779


\end{thebibliography}
\end{document}